\documentclass[10pt]{IEEEtran}
\makeatletter
\def\ps@headings{%
	\def\@oddhead{\mbox{}\scriptsize\rightmark \hfil \thepage}%
	\def\@evenhead{\scriptsize\thepage \hfil \leftmark\mbox{}}%
	\def\@oddfoot{}%
	\def\@evenfoot{}}
\makeatother
\usepackage[english]{babel}
\usepackage{lipsum}
\usepackage{indentfirst}

\usepackage{algorithm}
\usepackage{graphicx}
\usepackage{multicol}
\usepackage{mathtools}
\usepackage{balance}
\usepackage{commath}
\usepackage{bbding}
\usepackage{amsmath}
\usepackage{algpseudocode}
\usepackage{float}
\usepackage{graphicx}
\usepackage{subcaption}
\usepackage{bm}
\usepackage{amssymb}
\usepackage[font={small}]{caption}
\usepackage{array}
\usepackage{verbatim}
\usepackage[dvipsnames]{xcolor}
\graphicspath{ {./figures/} }

\newcommand*{\affmark}[1][*]{\textsuperscript{#1}}

\begin{document}
	
	\title{Toward a real-time TCP SYN Flood DDoS mitigation using Adaptive Neuro-Fuzzy classifier and SDN Assistance in Fog Computing}
	\vspace{-5mm}
	\author{\IEEEauthorblockN{Radjaa Bensaid\affmark[1], Nabila Labraoui\affmark[2], Ado Adamou Abba Ari\affmark[3], Leandros Maglaras\affmark[4], Hafida Saidi\affmark[1], Ahmed Mahmoud Abdu Lwahhab\affmark[5],and Sihem Benfriha\affmark[1]}\\

  {\affmark[1]STIC Laboratory,  Abou Bekr Belkaid Tlemcen University , Algeria}\\
  {\affmark[2]LRI Laboratory,   Abou Bekr Belkaid Tlemcen University, Algeria}\\ 
  {\affmark[3]DAVID Laboratory, Paris-Saclay University,France }\\
   {\affmark[4]School of Computing, Edinburgh Napier University, Edinburgh, UK} --- Corresponding author\\
   {\affmark[5]Department of Electronics and Communications, Dakahlia Mansoura University, Egypt}\\}	
		\vspace{-5mm}

\maketitle


\begin{abstract}
The growth of the Internet of Things (IoT) has recently impacted our daily lives in many ways. As a result, a massive volume of data is generated and needs to be processed in a short period of time. Therefore, the combination of computing models such as cloud computing is necessary. The main disadvantage of the cloud platform is its high latency due to the centralized mainframe. Fortunately, a distributed paradigm known as fog computing has emerged to overcome this problem, offering cloud services with low latency and high-access bandwidth to support many IoT application scenarios. However, Attacks against fog servers can take many forms, such as Distributed Denial of Service (DDoS) attacks that severely affect the reliability and availability of fog services. To address these challenges, we propose mitigation of Fog computing-based SYN Flood DDoS attacks using an Adaptive Neuro-Fuzzy Inference System (ANFIS) and Software Defined Networking (SDN) Assistance (FASA). The simulation results show that FASA system outperforms other algorithms in terms of accuracy, precision, recall, and F1-score. This shows how crucial our system is for detecting and mitigating TCP SYN floods DDoS attacks.
\end{abstract}

Keyword Fog computing, Service availability, TCP SYN flood DDoS attack, ANFIS, SDN.

\section{Introduction}
\vspace{1ex}


The growing number of connected objects, from millions to billions in various fields, is leading to an explosion in the amount of data. These huge volumes of data cause a lack of latency and make real-time analysis complex and difficult. To solve these issues, the deployment of computing models such as cloud and fog computing is crucial \cite{ar1}. Technologies of cloud computing enable an extremely powerful computer resource over the network. Nevertheless, due to several concerns about data privacy and security, attaching more diverse types of objects immediately to the cloud is extremely difficult, as well as network latency difficulties \cite{ar2, ar3}. Therefore, the need to introduce a new paradigm is necessary to solve these problems. 

Recently, fog computing has emerged to expand the cloud computing paradigm from the core to the network's periphery. The purpose of fog computing is to bring computer capabilities closer to IoT devices, offering real-time processing with low latency \cite{ar4}. Aside from this, fog computing also provides mobility support, location awareness, and decentralized infrastructure. Fog computing has a local data storage infrastructure, which makes it more secure than cloud computing. Despite this, IoT devices are limited in terms of storage capacity and battery life. Thus, they can be easily hacked, destroyed, or stolen and fog computing may become unavailable and unable to handle normal user requests. Therefore, it is necessary to apply security mechanisms to identify and block unauthorized requests on network systems. However, fog computing is still susceptible to various security and privacy gaps. It can be a point of vulnerability, and it is easily overwhelmed by a massive number of malicious requests, primarily intended for Distributed Denial of Service (DDoS) attacks \cite{ar5}.  DDoS attacks can be divided into two types depending on the protocol level addressed. The first one is known as "network-level flooding," when TCP, UDP, ICMP, and DNS packets are used to overload intended clients' network capabilities and resources. Whereas, the second protocol level is referred to as "application-level DDoS flooding" which is typically done on an HTTP webpage when attacks are launched to deplete server resources such as sockets, CPU, ports, memory, databases, and input/output bandwidth \cite{ar6}.
Regarding the rapid growth and the harm caused by DDoS attacks, several kinds of research have been conducted on these attacks, and various approaches have been presented in the literature to prevent these attacks using fog computing \cite{ar7}\cite{ar8}. Most of them proposed a defensive fog computing that operates as a filtering layer among the user layer and the cloud computing layer. However, these defensive approaches miss DDoS detection mechanisms, and detailed computation is not discussed. Also, they do not identify any infrastructure to protect fog computing which is particularly susceptible to DDoS attacks and may disrupt network services.\\
For this purpose, proposing Software Defined Networking ( SDN) technology-based solutions could bring an innovative framework to deal efficiently with this insidious attack. SDN enables us to define logic control and instructs the forwarding plane to act appropriately by isolating the control and data planes. This programmability provides more control over network traffic, which wasn’t conceivable before the development of SDN \cite{ar61}. \\
Considerable research has been done within SDN-based IoT-fog networks using task scheduling techniques like Threshold Random Walk with Credit-Based connection (TRW-CB) and rate limiting. These techniques are deployed for detecting anomalies and mitigating DDoS attacks, which effectively reduces average response times \cite{ar59}. However, it can result in excessive CPU and RAM consumption. This scheduling-based approach only focuses on secure scheduling periods, leaving the network vulnerable during idle times when no tasks are scheduled \cite{ar60}. Moreover, the previous approach incorporates both fuzzy logic and multi-objective particle swarm optimization. Nevertheless, as the number of variables and rules increases, designing and fine-tuning the fuzzy logic system can become highly complex. \\
Recently, machine and deep learning algorithms have gained attention for their effectiveness in detecting DDoS attacks by analyzing data patterns \cite{ar9}\cite{ar10}. Hence, merging fuzzy systems with neural networks combines the benefits of neural learning with the interpretability offered by fuzzy systems. An Adaptive Neuro-Fuzzy Inference System (ANFIS), empowers fuzzy systems to acquire knowledge from data. This synergy enhances fuzzy systems through neural networks. ANFIS's hybrid approach facilitates adaptability to diverse attack patterns and network conditions.\\
Moreover, employing various approaches, such as the Neuro-Fuzzy classifier on the KDD CUP99 dataset \cite{ar11}, has been a common practice. This dataset includes numerous recognized attack variations and has traditionally been utilized in intrusion detection. Nevertheless, the KDD CUP99 dataset is now regarded as outdated, as it presents several unresolved issues that fail to meet the updated criteria for DDoS identification \cite{ar12}.
In our work, we focus exclusively on the TCP SYN flood attack. Since it is the most effective DDoS attack in fog computing. In order to exhaust the system’s resources or overwhelm the target server, the attackers typically infect several devices that behave as bots and synchronize suspicious traffic or requests, leading to an incomplete three-way handshake procedure \cite{ar13}. consequently, Legitimate users cannot reach the desired fog server.
 In this paper, we suggest a novel Fog computing-based SYN Flood DDoS attack detection and mitigation using an Adaptive Neuro-Fuzzy Inference System (ANFIS) and SDN Assistance (FASA). Compared to previous works, FASA utilizes the ANFIS model for network traffic classification, incorporates SDN support to enable real-time mitigation, and relies on the newly released CIC-DDoS2019 dataset. The proposed model demonstrates exceptional performance across multiple metrics, including accuracy, precision, recall, and F1- score. Additionally, it exhibits a notably low rate of false positives. In brief, our significant contributions are outlined as follows: 
\begin{enumerate}
\item We propose a novel model FASA to detect and mitigate a SYN Flood DDoS attack in fog computing using SDN assistance.
\item We implement the ANFIS model to self-train the fog servers and make the difference between normal and malicious packets.
\item The ANFIS model is implemented at the SDN controller and deployed at the fog server, using a dataset captured from the SDN environment. Its main objective is to allow benign packets access while rejecting malicious ones to release a secure and dependable SDN controller that ensures fog service availability.
\item The proposed evaluation method uses both the newly released dataset CIC-DDoS2019 and the SDN dataset. It is experimentally analyzed from the data availability and the algorithm operating efficiency and it can improve the performance
\end{enumerate}

The paper is structured in the following manner:  Section 2 examines and discusses previous works to tackle the issue of DDoS attacks. Section 3 contains background knowledge. In Section 4, we formally define the proposed model, and in Section 5, we introduce our proposed framework. followed by the evaluation outcomes and discussion in section 6. Finally, the conclusion is given in Section 7.

\section{Related Work}
\vspace{1ex}
\parskip 0pt
In this section, we have provided an extensive overview of DDoS attacks. especially TCP SYN flood attack detection. Besides, these works have been grouped into three sections. The initial represents statistical methods. The second and third ones highlight a few works based on Machine/Deep Learning (ML/DL) algorithms. 

\subsection{Statistical methods}
Statistical methods constantly evaluate user/network activities to identify abnormalities\cite{ar14}. Hence, due to their capacity to analyze the behavior of data packets, they are commonly utilized in DDoS attack detection systems. If data flow does not match with some test statistics and measures, it is thought to be illegal.
Ahalawat et al. \cite{ar15} suggested a detection method for DDoS attacks based on Renyi entropy. and a mitigation solution for SDN based on the packet drop approach, using several probability distributions. They can examine network traffic fluctuations. However, the necessity to set an optimal detection threshold is a typical limitation of various entropy-based approaches. Hoque et al. \cite{ar16} presented a novel correlation measure using standard deviation and mean to detect DDoS attacks, The traffic is then classified as attack traffic or normal by comparing the collected traffic to the profiled traffic. However, the suggested metric's use in identifying low-rate attempts is unclear. a DDoS detection-based multivariate correlation analysis was discussed by Jin et al. \cite{ar17} in their work, and they provided a covariance analysis method for recognizing SYN flood attacks. The experimental results demonstrate that this technology accurately and efficiently detects DDoS attack traffic in networks of varied levels of intensity. However, using the correlation approaches consumes a lot of processing in real-time to detect DDoS attacks. As a result, they are unable to operate in real-time.
A novel framework was suggested by Bhushan et al. \cite{ar18} using fog for detecting DDoS attacks even before they reach the cloud by using an efficient resource provisioning algorithm to service cloud requests through intermediate fog servers. Furthermore, an entropy DDoS detection method and mitigation system designed for Cloud Computing environment using SDN has been proposed by Tsai et al. \cite{ar19}.  An entropy-based DDoS detection approach was implemented to protect the virtual machines and controller from malicious attacks. As a result, the detection rate is significantly affected by the threshold value.
Javanmardi et al. \cite{ar59} proposed FUPE, a security-driven task scheduling algorithm for SDN-based IoT-Fog networks. FUPE uses fuzzy logic and multi-objective particle swarm optimization to assign tasks to fog nodes balancing security and efficiency objectives. However, managing and interpreting extensive rule sets pose challenges in maintaining and validating the fuzzy logic framework. Nonetheless, Multi-objective optimization with PSO requires parameter tuning and could be computationally intensive, particularly in large-scale environments. 
Furthermore, it incorporates techniques like Threshold Random Walk with Credit-Based connection (TRWCB) and rate limiting to detect malicious nodes and utilizes the SDN controller for mitigation by blocking attackers, ultimately leading to a reduction in average response times. Nevertheless, this approach may lead to elevated CPU and RAM usage. FUPE exclusively identifies and addresses anomalies during the scheduling phase, leaving the network susceptible to threats in the absence of scheduling requests \cite{ar60}.

\subsection{Machine Learning methods (ML)}
Machine learning-based methods are used to identify DDoS attacks such as Decision Trees, Deep Learning, Support Vector Machine (SVM), K Means Clustering, and so on \cite{ar20}. These methods might be unsupervised machine learning (label for training is not required) or supervised machine learning (require a label for training normal/malicious) algorithms. Moreover, the dataset, which contains numerous network and traffic features, is used to train and learn automatically how to recognize suspicious behavior patterns. 
Rajagopal et al. in \cite{ar21} provided a meta-classification strategy that integrates many classifiers for both binary and multiclass classification. Decision jungle serves as the meta learner, combining numerous learners to obtain the best prediction performance. This proposed method has a precision of $99\%$.  
Tuan et al. \cite{ar22} idea were about proposing a novel TCP-SYN flood attack mitigation by tracing back IP sources of attack in SDN networks using K-Nearest Neighbors (KNN) machine learning based on SDN. The testbed's experimental findings reveal that $97\%$ of attack flows are identified and blocked.
Priyadarshini et al. \cite{ar23} demonstrated a new source-based DDoS mitigation approach, in order to prevent these attacks in both fog and cloud computing environments. It deploys the defender module that presents at the SDN controller which is based on machine learning (SVM, KNN, and Naive Bayes algorithm).  However, the classical ML techniques can't handle the amount of data. 
However, the "real world" application of classical ML algorithms is limited due to network attack issues. In addition, these approaches need a lot of time to learn, thus they can't be used in real-time.

\subsection{Deep Learning methods (DL)}
In recent studies, there has been a particular emphasis on evaluating the performance of DL models in DDoS detection. This is primarily due to their ability to effectively analyze large volumes of data and identify complex patterns within it.
Assis et al. \cite{ar24} proposed a near real-time solution by applying Convolutional Neural Networks (CNN) to cover and defend victims' servers from DDoS attacks at the end source, The detection model reached a precision rate above $ 95.4\%$. Novaes et al. \cite{ar25} employed the Generative Adversarial Network (GAN) architecture to mitigate the damage of DDoS attacks on SDNs. For experiment assessments, the accuracy obtained using the published datasets namely, CIC-DDoS2019 and the emulation was about $ 94.38\%$. The authors compared the GAN framework's findings against those of other deep learning algorithms, such as LSTM, CNN, and MLP. The authors of \cite{ar26} employed a variety of Machine Learning (ML) algorithms to identify low-rate DDoS attacks. They found that the Multi-Layer Perceptron (MLP) performs the best among the assessed algorithms, with a detection rate of up to $95\%$. Other ML models, such as Random Tree, Random Forest, and Support Vector Machines, have shown useful in detecting and mitigating DDoS attacks. Deep learning has already been used to identify SYN flood attacks by Brun et al. \cite{ar27}, in which a Random Neural Network was built to classify and differentiate whether the packet is normal traffic or SYN attacks. Evmorfos et al. \cite{ar28} use a Random Neural Network for identifying typical SYN attacks on Internet-connected equipment including edge devices and gateways, and fog servers, with limited processing capability. Devi et al. \cite{ar29} presented an intrusion detection system (IDS) approach based on the SUGENO-based fuzzy inference system ANFIS to identify security concerns on relay nodes in a 5G wireless network. The model was tested and trained using the KDD Cup 99 datasets. Boroujerdi et al. \cite{ar30} developed a novel ensemble of Sugeno-type adaptive neuro-fuzzy classifiers to identify DDoS attacks based on the Marliboost boosting approach. It was tested on the NSL-KDD dataset. However, the data in the NSL-KDD or KDD cup 99 datasets were considered unsuitable for the new requirement of a DDoS attack since it comprises packet traces rather than flows, implying that the DDoS detection methods may become computationally difficult as the network expands in size. As a consequence, there have been various studies published in recent years on how to identify DDoS attacks, particularly TCP SYN flood using Machine and Deep Learning. However, few of them have addressed using ANFIS to detect such attacks in fog computing based on SDN technology.

In order to address the limitations of the previous studies, in this paper, we propose an ANFIS classifier, implemented in the SDN controller to classify network traffic, and deployed at the fog server using the recently published dataset CIC\-DDoS2019. The inclusion of various types of DDoS attacks in this dataset bridges the gaps found in previous databases. Additionally, we employ the ANFIS using the SDN dataset for real-time mitigation. 

\section{Background knowledge}
\label{Background knowledge}
This section highlights the required context for our proposed model. First, we give an overview of DDoS attacks and the different methods used for detection. Then, we introduce the Adaptive Network-based Fuzzy Inference System (ANFIS) detection algorithm. Finally, we present the Software Defined Networking (SDN) technology.

\subsection{DDoS attacks and fog computing}
The DDoS attack is a highly progressed type of DoS attack. It differs from other attacks in that it may be deployed in a "distributed" manner. A DDoS attack's primary purpose is to inflict harm on a target for personal reasons, financial gain, or popularity. \cite{ar1}. It is an attack based on availability and aims at making the victim system inaccessible to authorized users \cite{ar31}. Moreover, it is done by a combination of a huge amount of hacked and dispersed devices known as bots or zombie devices that have been infected with malicious malware or compromised by an attacker \cite{ar32}. Hence, an attacker centrally controls and coordinates these machines to launch an attack on the target machine \cite{ar33}.

\subsubsection{Types of DDoS attacks on fog computing}
Several DDoS attack types are used to bring down the functionality or availability of network services on fog computing [34], as illustrated in Figure 1.

\begin{enumerate}
\item[a.] \textbf{Application-Bug Level DDoS} \\
These sorts of attacks, like HTTP POST and HTTP PRAGMA, deplete the application system, causing it to fail or temporarily close down. 

\item[b.]\textbf{Infrastructural Level DDoS} \\
The key purpose of these threats is to exhaust network bandwidth, buffers, CPU, and storage, preventing legitimate users from using them. Thus, the only requirement for this attack is the victim’s IP address. It is categorized into two types: direct and reflector attacks.

\begin{itemize}
\item \textit{ Direct Attack} \\
This attack is carried out with the assistance of compromised devices or bots. It sends malicious queries to the target using bots in order to deplete its resources, bandwidth, and services, rendering them inaccessible to authenticated users. This attack can be further subdivided into network-layer and application-layer DDoS attacks.

\begin{itemize}
    \item Network Layer DDoS: This attack type employs various network and transport layer protocols, including TCP SYN, UDP, ICMP, among others.
    \item Application Layer DDoS: In this attack, HTTP flood traffic is adopted widely to exhaust the victim. this kind of vulnerability is difficult to detect, raising security issues.
\end{itemize}    
\item \textit{Reflector Attack} \\
In this attack, the IP address is spoofed and requests are delivered to a vast range of reflector hosts. Following the receipt of the requests, a response is provided in order to flood the target.
\end{itemize}
\end{enumerate}
\begin{figure}[h]
\centering
\includegraphics[width=0.5\textwidth]{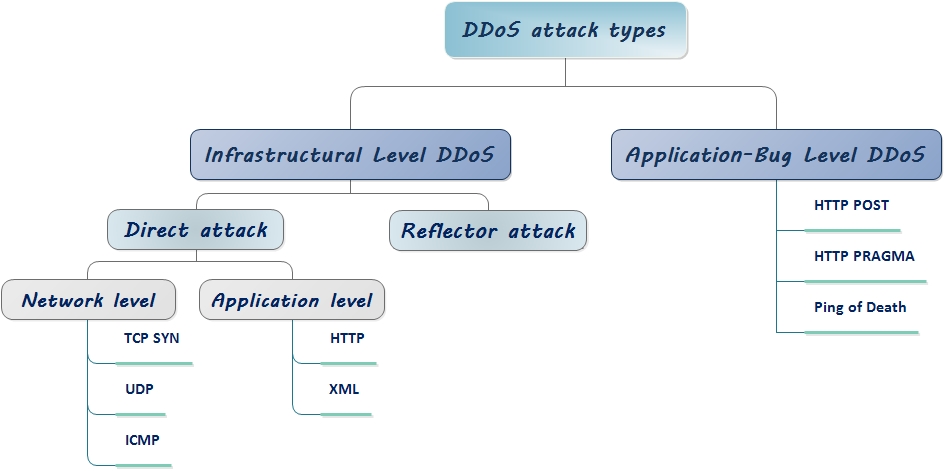}
\caption{Types of DDoS attack in fog computing}
\label{img1}
\end{figure}
\subsubsection{DDoS defense mechanisms} 
In this section, we discuss various defense mechanisms used for DDoS attack detection and mitigation for the security of fog computing \cite{ar34}. Nearer-to-edge devices, fog computing offers computing capabilities in the form of fog nodes. which creates a heavy load on network management. To address this issue, SDN technology can be implemented to guarantee the safety of fog computing in the following aspects:
\begin{itemize}
\item \textit Monitoring the network: If the network is monitored permanently and continuously, any suspicious data attempting to disrupt services may be recognized and rejected. As this is performed at fog nodes, legitimate users will have no difficulty accessing the services. 
\item \textit Priority-based and isolated traffic: It implies the process of prioritizing legal and illegitimate network traffic, hence requiring the use of shared knowledge resources such as CPU or I/O. As a result, SDN can reject damaging traffic by separating it through VLAN ID/tag.
\item \textit Access control mechanism for resources in the network: To prevent DDoS attacks, an effective access control system should be implemented.
\item \textit Shared network: The shared network is the crucial condition since anyone can access it, holding security at risk.
\end{itemize}
Additionally, two distinct assessments are used to identify DDoS defense mechanisms. The first classification divides the DDoS defense systems into the following four groups based on the activity carried out:
\begin{itemize}
\item \textit Intrusion Prevention,
\item \textit Intrusion Detection,
\item \textit Intrusion Tolerance and Mitigation, 
\item \textit Intrusion Response. 
\end{itemize}
Further, the second categorization mainly classify DDoS defenses into the following three groups based on where they are deployed:
\begin{itemize}
\item \textit Victim Network,
\item \textit Intermediate Network,
\item \textit Source Network.
\end{itemize}
\subsubsection{TCP SYN Flood attack } 
The SYN flooding attack is a specific type of DoS attack that targets hosts that operate TCP server processes, it became well-known in 1996 \cite{ar35}. The concept of the three-way handshake that initiates a TCP connection serves as the mainstay of this attack. \cite{ar36}. It exploits a TCP protocol process characteristic and may be used to restrict server functions from responding to normal user demand to establish new TCP connections. As a result, each service that connects and waits on a TCP socket is highly susceptible to TCP SYN flood attacks. Although several techniques to counteract SYN flood attacks may be found in modern operating systems and equipment.
\subsection{The Adaptive Network-based Fuzzy Inference System (ANFIS) detection algorithm}
ANFIS is a network model that combines a Sugeno-type fuzzy system with neural learning capability \cite{ar37}. Neuro-fuzzy systems are ways to learn fuzzy systems from data that use neural network-derived learning algorithms. Therefore, due to their learning capabilities, neural networks are an ideal choice for combining with fuzzy systems\cite{ar38} are used to automate or simplify the process of developing a fuzzy system for specific usage. The initial neuro-fuzzy techniques were primarily explored within the field of neuro-fuzzy control, although the approach is now broader because it is used in a number of domains, including control, data analysis, decision support, and so on. \cite{ar39}.  
ANFIS is based on two parameters (premise and consequent parameters) which are used to connect the fuzzy rules. Moreover, ANFIS is made up of five layers in total, as illustrated in Figure 2. The square nodes have parameters, whereas circular nodes do not.

The considered fuzzy inference system contains two inputs, y considered as non-linear parameters, and one output f. Also, each input variable is described by two linguistic terms: $A_1$ and $A_2$ for the variable x, and $B_1$ and $B_2$  for the variable y, respectively. \\
the following two \emph{IF-THEN} rules construct the Sugeno fuzzy model \cite{ar39}:

 \begin{itemize}
\item \textbf{Rule 1:} \text {If}\ x \text { is } $A_1$ $\wedge y$ \ \text { is } \ $B_1$, \ \text {then} \ $f_{1}=p_{1}x+q_{1} y+r_{1}$

\item \textbf{Rule 2:}\ \text {If}\ x \text { is } $A_2$ $\wedge y$ \ \text { is } \ $B_{2}$, \ \text {then} \ $f_{2}=p_{2}x+q_{2} y+r_{2}$
\end{itemize}

Where $p_{i}$, $q_{i}$, and $r_{i}$ {i=1,2},  correspond to the linear parameters of the conclusion part to be adjusted during the training.
\begin{figure}[h]
\centering
\includegraphics[width=0.5\textwidth]{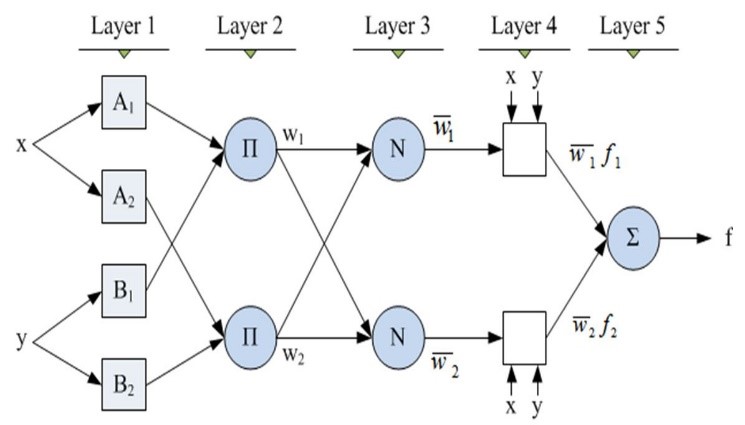}
\caption{ANFIS architecture layers}
\label{img2}
\end{figure}

\begin{itemize}
    \item \textbf{Layer 1:} O1, represents the membership function  µ  of a fuzzy set \(A_i (or B_i) \).
$$ O_{1,i}=\mu_{A_i} (x) \qquad i=1,2 $$ 
$$ O_{1,i}= \mu_{B_{i-2}} (y) \qquad i=3,4 $$  \\
Because of their smoothness and simple syntax, Gaussian membership functions are preferred approaches for defining fuzzy sets. The advantage of these curves is that they are smooth and nonzero at all locations. The Gaussian membership function is used in this study, it’s frequently used to reduce the uncertainty of real-world measurement and is represented by the equation (1) where c and $\sigma$  represent the mean and standard deviation respectively. Here c represents the center, and $\sigma$  represents the width. {a,c} are called premise parameters (non-linear).
\begin{equation}  
\mu_A (x)=ae^ \frac{-(x-c)^2}{(2\sigma)^2}       	 
\end{equation}
\item \textbf{Layer 2:} the fuzzification layer w determines the degree of membership function satisfaction of each input; the output is the product $\prod$ of all the entering signals, it is determined using the following equation (2):
\begin{equation}
O_(2,i)=w_i=\mu_A (x)\cdot \mu_B (x)  \qquad i=1,2 
\end{equation}                              
The output of every node shows the firing strength of a rule. The node function in this layer can be any other fuzzy AND T-norm operator, such as min.
\item \textbf{Layer 3:} the normalization layer, in which the i-th node determines the proportion of the firing strength of the   i-th rule to the  total firing strength of all rules, as demonstrated in the equation (3): 
\begin{equation}
O_{3,i}=\bar{w_{i}}=\frac{w_{i}}{w_{1}+w_{2}} 
\end{equation}
The outputs of this layer are referred to as normalized firing strengths.
\item \textbf{Layer 4:} In the defuzzification layer, parameters are named consequent parameters. Each node has a function where $\bar{w_{i}}$ is a normalized firing strength from layer 3 and {$p_i$,$q_i$,$r_i$} are the set of linear node parameters and are defined as consequent parameters of this node and $f_i$ denotes the output of the rule, as shown in equation (4):
\begin{equation}
O_{4,i}=\bar{w_{i}}f_{i}= \bar{w_{i}}(p_{i}x+q_{i}y+r_{i})
\end{equation}\\
\item \textbf{Layer 5:} in this layer, the single node adds up all of the incoming signals to compute the overall output, as demonstrated in equation(5):\\
\begin{equation}
O_{5,i}= overalloutput= \sum_{i= 0}^{n} \bar{w}_{i}f_{i} =\frac{\sum {w}_{i}f_{i}}{\sum {w}_{i}} 
\end{equation}
\end{itemize}                                      
An adaptive network's nodes are related to parameters that may affect the final output. To adapt the parameters in an adaptive network, ANFIS typically uses a hybrid learning algorithm, that associates gradient descent and the least square approach \cite{ar40}.
The hybrid algorithm comprises a forward pass and a backward pass. To optimize the consequent parameters, the least squares method (forward pass) is used; node outputs are passed forward until Layer 4, and the least squares determine the consequent parameters. In our work, For optimizing the premise parameters, the ADAM method \cite{ar41} is employed during the backward pass. Error signals are propagated backward, and the premise parameters are updated using ADAM. This hybrid learning approach offers faster convergence by reducing the search space dimensions compared to the original backpropagation method. \cite{ar42}. It has been demonstrated that this hybrid algorithm is extremely effective in training ANFIS systems \cite{ar43}.
The ANFIS training technique begins by defining the number of fuzzy sets, the number of sets of each input variable, as well as the shape of their membership function. The primary goal of ANFIS is to improve input-output data sets and a learning mechanism to enhance the parameters of a comparable fuzzy logic system. The difference between the intended and actual outputs is minimized as much as feasible during parameter optimization.
\subsection{Software Defined Networking (SDN)}
SDN is a network paradigm that enables users to directly manage network resources by orchestrating, controlling, and using software applications \cite{ar44}. Moreover, the control and data planes are divided by SDN. making it most commonly used to improve network efficiency. When the data plane forwards packets from one location to another, the control plane determines whether or not the packets should propagate through the network. 
Thus, SDN is formed by the combination of a controller and switches, these switches follow the forwarding rules that are defined by the controller, which can dynamically manage network flows and implement different configurations based on network circumstances. The three fundamental layers of SDN architecture are (i) The application layer which contains the general network functions including intrusion detection systems, firewalls, and security applications. (ii) The control layer which is the centralized software controller that serves as the SDN's brain.  The network policies and traffic flows are managed by this controller. (iii) The infrastructure layer contains a variety of networking equipment, including switches and routers \cite{ar45}, as shown in Figure 3. The communication between the controllers and switches is outlined throughout the OpenFlow protocol \cite{ar46}, which serves as the communication standard for SDN networks. It is referred to as SDN networks' southbound communication. The controller can deal with open flow switches (OF-switch) with existing flow tables by using an open flow protocol. When a packet's flow entry is found in the OF-switch's table, the packet is forwarded in the usual manner; otherwise, The controller receives it for additional evaluation. Thus, SDN controllers with OpenFlow-enabled switches are widely used for SDN networking. They are especially suitable for light traffic communication and control.
\begin{figure}[ht]
\centering
\includegraphics[width=0.5\textwidth]{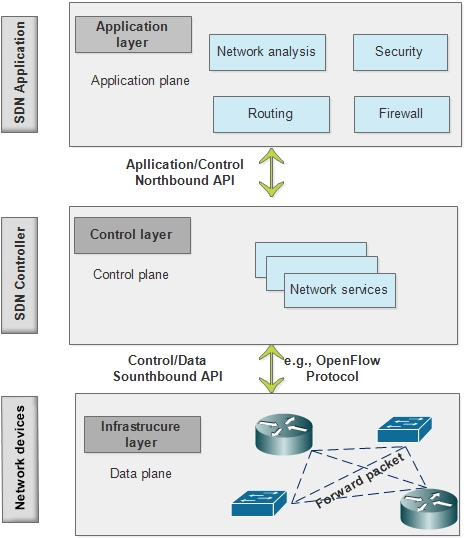}
\caption{Software Defined Networking}
\end{figure}

\subsection{System Model}
This study aims to develop a distributed FASA framework to mitigate SYN flood attacks in the network environment by recognizing and avoiding attacks close to the attacking sources. To enable quicker and more accurate attack detection using the ANFIS model, fog computing is suitable for deploying SDN for mitigating SYN flood attacks by assigning compute power near the operation process and spreading the burden in the system through a FASA mitigation scheme. In this section, we first outline SYN flood DDoS attacks in fog computing. Then, we discuss the FASA network architecture. 

\subsection{SYN Flood DDoS attack} 
As shown in Figure 4, when a standard TCP three-way handshake has initiated, the End User (EU) transmits the SYN packet to the fog server. Then, the fog server responds with an SYN/ACK packet. Next, the EU should send an ACK packet to the fog server. So, when all of these processes are completed, the connection is established \cite{ar47}. However, the main drawback of TCP connections is the inability to maintain half-open connections. The fog server is in a half-open connection state because it is standing in line for the EU's reply to acknowledge the three-way handshake. Furthermore, IoT devices have limited computation, storage capacity, and short battery life, and they can easily compromise, damaged, or kidnapped. Therefore, due to the aforementioned limitations, an attacker may simply hack IoT devices and utilize them as botnets to generate and send excessive SYN request packets with a fake source IP address to fog servers. As a result, the ACK packet will never reach the fog server which is in the open port state waiting for the ACK packet. Moreover, the SYN/ACK packets are transmitted to the faked host, and the three-way handshake procedure will never be completed. Also, the connection registration is kept in the connection delay buffer till time expires, preventing legitimate users from accessing the services \cite{ar48}. 

\begin{figure}[h]
\centering
\includegraphics[width=0.52\textwidth]{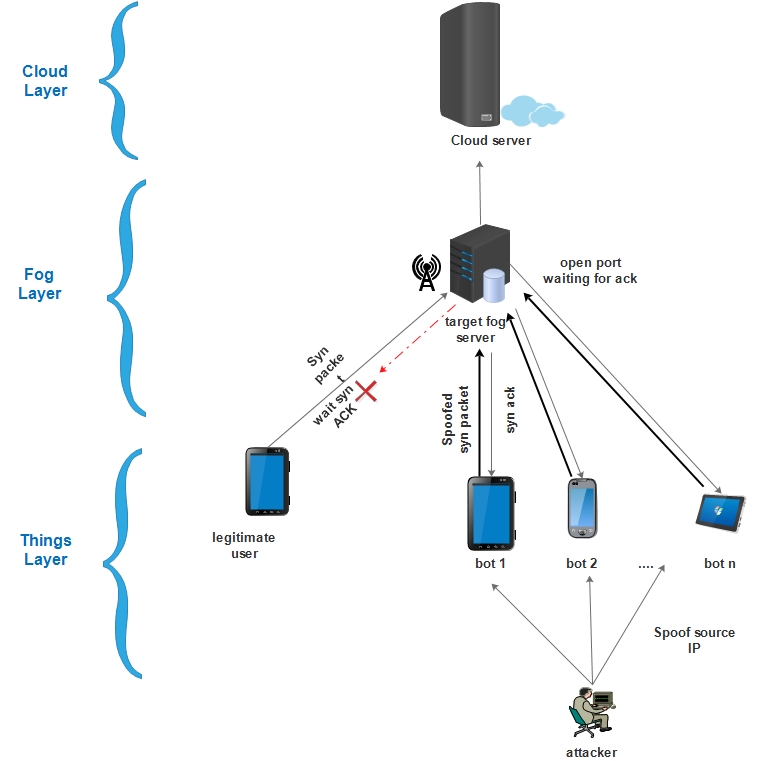}
\caption{SYN flood DDoS attack in fog computing}
\end{figure}

\subsection{ FASA Network Architecture} 
To effectively deal with the SYN flood DDoS attack concerns in the network systems, attack prevention must be built in fog computing based on SDN. Indeed, in this paper, we propose a novel distributed fog defensive system for SYN flood DDoS attacks using ANFIS and SDN Assistance (FASA). The FASA architecture has three layers, the cloud layer, the SDN-based fog (SDFN) Layer, and the things layer, as shown in Figure 5.
 \begin{enumerate}
\item[a.] \textbf{Cloud layer}: cloud computing, as a computing model, define a method of managing a pool of configurable computing resources, offers elastic, on-demand services, and has access to the system anywhere and at any time. Therefore, users can use resources according to their demands. The salient features provided by cloud technology are immediate flexibility and measurable services. \cite{ar1}
SDN and cloud technology can be combined to automate and cloud applications provisioning must be completely integrated with the network.  Hence, in the FASA system, cloud computing refers to the application plane which consists of many useful applications that communicate with the controller to abstract a logically centralized controller to make coordinated decisions.

\item[b.] \textbf{SDN-based Fog Network layer (SDFN)}: This layer combines the fog computing and SDN paradigm to identify and behave against DDoS attacks. With recent advances in SDN, it opens up new opportunities for providing intelligence within networks. The benefits of SDN, including logically centralized control, software-based traffic analysis, an entire network view, and flexible forwarding rule updates, help to improve and facilitate machine learning applications \cite{ar49}. Therefore, the SDFN layer provides new trends of DDoS attacks in fog computing environments using SDN. This layer is formed of two sub-layers, SDFN-server and SDFN-node.
\begin{itemize}
\item \textbf{SDFN-server:} This sub-layer refers to the control plane deployed at fog servers where an intelligent ANFIS classifier is integrated into the control network to classify traffic flows decision and consequently policies are managed to depend on its decisions. 
Moreover, the SDFN server communicates with the cloud layer (application) via the northbound interface and with the SDFN-node layer via the southbound interface.
\item \textbf{SDFN-node:} This sub-layer refers to the data plane of physical equipment in the network such as switches and routers. It forwards the network traffic to their destinations using the OpenFlow protocol.
\end{itemize}
\item[c.] \textbf{Things layer:} This layer serves the purpose of sensing, collecting, and uploading data from wireless sensors and end-users to fog computing. The transmitted packet can be classified as either benign or malicious.
 \end{enumerate}

The following assumptions are made in order to better explain the SYN flood DDoS attack identification and defense framework: 
\begin{itemize}
\item \textit The SDN-based Fog Network server (SDFN-server) is susceptible to being compromised, 
\item \textit DDoS attacks are TCP SYN flood attacks against SDFN-servers.
\item \textit The SDN controller and the switch are not compromised. 
\item \textit IoT devices can be hacked.
\end{itemize}
 \begin{figure}[h]
\centering
\includegraphics[width=0.5\textwidth]{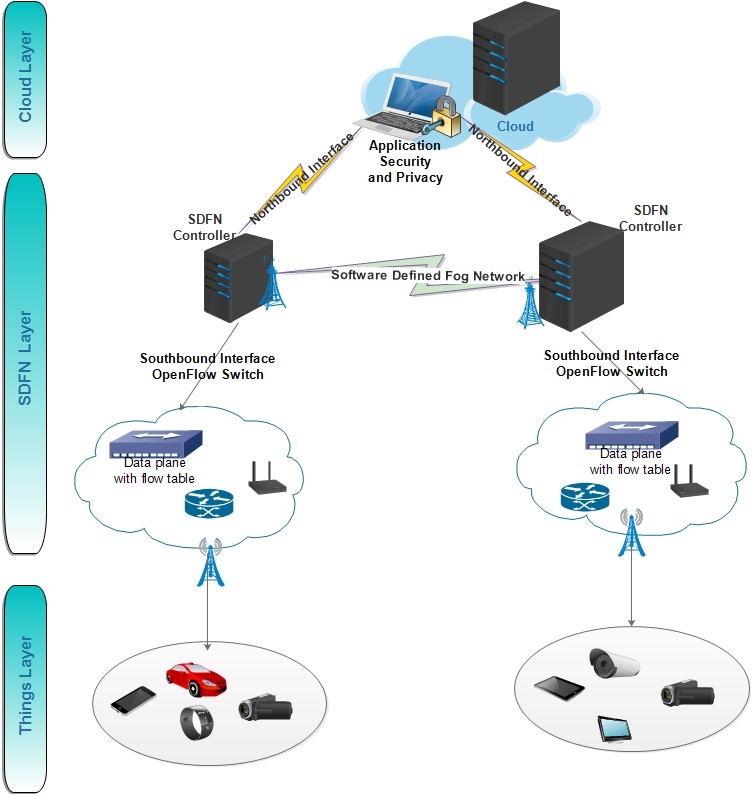}
\caption{Network model: SDN-based Fog Network (SDFN) }
\label{img5}
\end{figure} 


\section{Proposed FASA framework}
SYN flood DDoS attacks can instantly bring down a network and it is difficult to detect them since they can be carried out in a very short time. Therefore, detecting and mitigating such attacks is critical. A detection approach for such threats is needed in fog computing to filter and block the malicious requests before the attack produces a negative impact on the fog services. Consequently, our FASA framework can be used to identify and immediately mitigate SYN flood attacks in real-time into fog computing, as illustrated in Figure 6.  
\begin{figure}[h]
\centering
\includegraphics[width=0.48\textwidth, height=0.3\textheight]{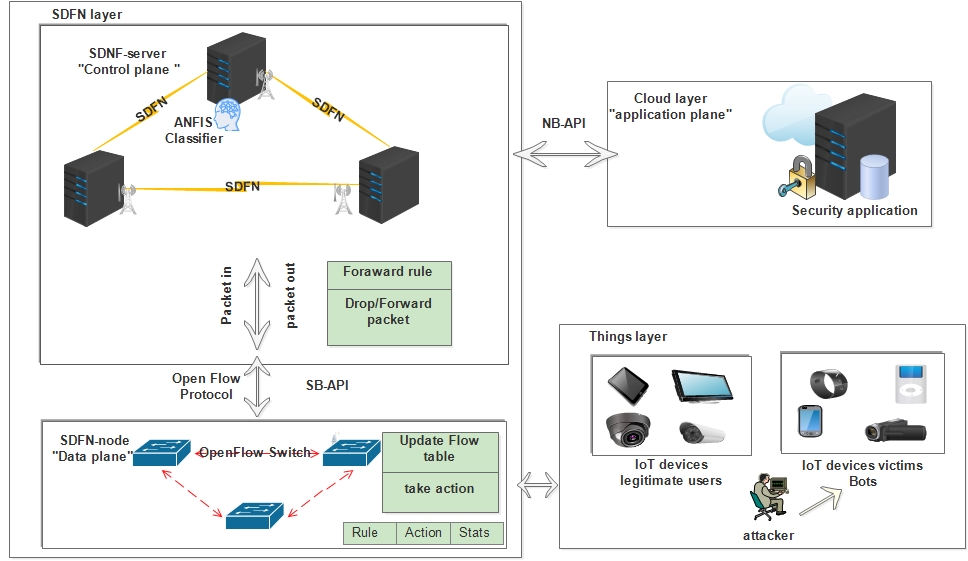}

\caption{The proposed FASA framework }
\label{img6}
\end{figure}

\subsection{The detection process}
FASA is based on the ANFIS model and SDN network to guarantee service availability in the fog network. To attain recognition and detection purposes, a fog layer is established among both the cloud layer and the Things layer. Thus, the recognition techniques deployed on the fog layer can handle and process malicious traffic. Also, the SDN controller deployed on the fog layer controls packets arriving from every system node to enhance security and network management. 
Additionally, the SDFN-server is prior trained with ANFIS algorithms and tested using two different datasets, CIC-DDoS2019 and SDN dataset. After a successful data pre-processing step, the most important features will be extracted. Then, these features will be divided into training data and testing data to self-train the SDFN-server to identify the SYN flood attack. Once that is done, the ANFIS model will be able to determine whether an incoming packet is legitimate or not. then, the controller's decision based on that.  as presented in the flowchart of Figure 7.
\begin{figure}[h]
\centering
\includegraphics[width=0.5\textwidth]{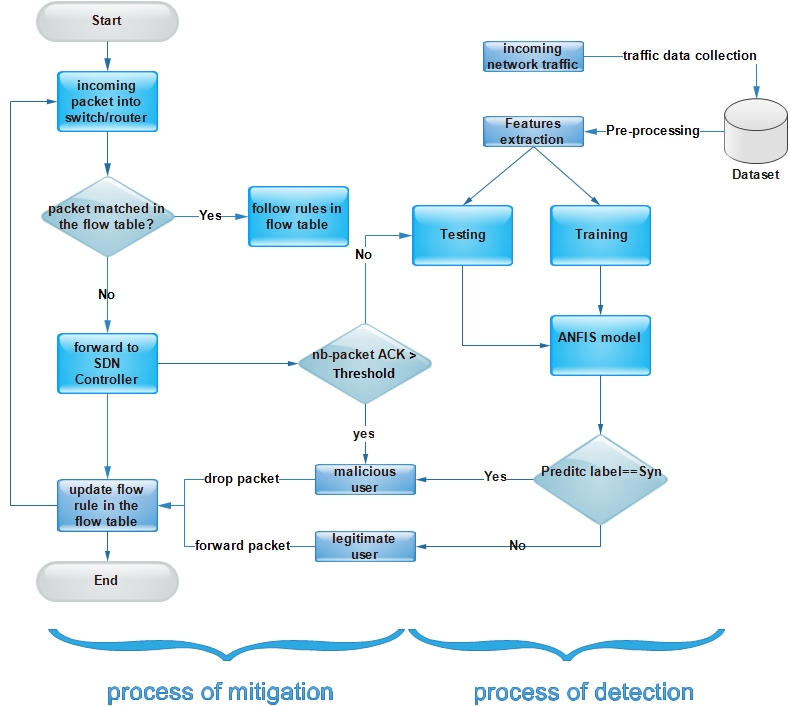}
\caption{Flowchart of the proposed model FASA}
\label{img7}
\end{figure}

\subsection{The mitigation process}
SDN simplifies the implementation of complex mitigation models. When an OpenFlow switch gets a packet, it compares it to the matching rule in its flow table and decides whether to act by forwarding packets to the destination according to the found rule or seek assistance from the controller if the rule is not matched. In addition, the OpenFlow switch initiates this request through the SB-API by the OpenFlow agent in the switch, as demonstrated in the flowchart of Figure 7.
 Although, the attack may be identified by determining a threshold value, which is the maximum value of serving capacity defined by the availability of computational resources. If the number of service requests exceeds the limit, a malicious packet is sent out \cite{ar50}. Otherwise, if it’s less than the Threshold capacity, it will pass through the ANFIS classifier for prediction in the fog server. Therefore, the real-time mitigation phase is started when the ANFIS model detects an SYN flood packet. This phase aims to perform defensive functions to limit the damage caused by an exploit. So, the packet passes through the OpenFlow protocol which takes action by executing the updated rule in the flow table whether it is a legitimate user to allow access. Otherwise, the controller looks for the most often occurring source address Mac with different source address IPs and uses it to determine the infected port number.  By correlating the identified Mac address with the corresponding port on the switch, the controller determines the port through which the attack traffic is entering the network. To prevent further damage, the controller instructs the OF switches to drop all packets obtained from the host associated with the identified Mac address. Then, The controller also directs the switch to block traffic on the specific port associated with the infected host, effectively preventing any communication through that port. Next, the controller updates the flow table of the switch to modify the rules related to receiving or forwarding packets to the identified port. This ensures that any packets destined for that port are dropped or redirected to mitigate the attack. \\ 
 As a result, TCP SYN flooding attacks may be identified and prevented by instantly blocking the switch port that is connected directly to the attacker's host. 

 \begin{algorithm}
\begin{algorithmic}
\caption{FASA framework process}\label{alg:cap}
\State {input: \textbf{incoming packet of traffic flow to the switch} }
   \State {output: \textbf{response with flow classification and decision}}
\If{packet matched in the flow table} 
    \State Apply the rule in the flow table;
\Else \State Forward packet to SDFN-server;
       \State Apply ANFIS classifier;
    \If{flow classified as malicious packet}
    \State Retrieve the Mac address of the attacker;
   \State Update rule table in flow table with a malicious user;
   \State  Make a decision: 
   \State Drop the packets with this source Mac address;
   \State Block  the infected switch port;
   \Else \State Update rule table in SDN with the legitimate user;
 \State Make decision: Forward the packet to destination;
\EndIf
\EndIf 
\end{algorithmic}
\end{algorithm}

\section{Experiments and results}
\subsection{Experimental setup}
In this part, we will go over the various tools that were used to build up the experimental setup for detecting SYN flood attacks in the simulated SDN and fog computing environments, using Wireshark \cite{ar51} to capture and analyze network traffic in real-time. The entire experiment is carried out on Windows 10 OS with an Intel i3 processor and 8GB of RAM.
To emulate the network behavior, the SDN Mininet network emulator \cite{ar52} was used, with the Ryu controller \cite{ar53}. Ryu is an open-source platform, that provides transparency and flexibility, enabling customization and extension of functionalities. Its Python-based architecture promotes accessibility and ease of development, facilitating rapid implementation of SDN applications. Additionally, support for multiple protocols, including OpenFlow, ensures seamless communication with diverse network devices. Ryu's compatibility with various networking technologies and hardware makes it suitable for heterogeneous infrastructures, rendering it particularly well-suited for this research \cite{ar62}.\\
For training and testing our ANFIS model, the Python programming language has been used with libraries for deep learning Keras \cite{ar54}, and TensorFlow \cite{ar55}. Additionally, to prevent overfitting, the stratified K-Fold cross-validation \cite{ar56} was also employed in the ANFIS algorithm. Due to the fact that the Stratified k-fold cross-validation guarantees that each fold has a class distribution that is identical to the original dataset, resulting in a more accurate and reliable model assessment.
, along with Binary Crossentropy, a classic loss function used in binary classification. Also, we set the default Keras learning rate to 0.001. Furthermore, Adam optimizer\cite{ar41} was selected, as an adaptive algorithm for optimizing learning rates in neural network models.
Moreover, by using two different scenarios in this study, we examine the performance and efficiency of the FASA system.
\begin{itemize}
\item \textit{Scenario 1}: Evaluate the performance of the FASA system by employing the SDN environment.
\item \textit{Scenario 2}: Evaluate the performance of the FASA system by using the public dataset CIC-DDoS 2019 \cite{ar57}.
 \end{itemize}

\subsection{Experimental analysis}
In our next subsection, we discuss each test scenario and provide the studies' results.
 \begin{enumerate}
\item[a.] \textbf{Scenario 1}
In our experiment, the Mininet network emulator [51] was used to design virtual network topologies consisting of controllers, hosts, links, and switches. Therefore, to run Mininet and Ryu controllers \cite{ar53}, we have used two virtual machines based on the Linux operating system.  Ryu controller is based on a Python program and supports several network management protocols such as OpenFlow switches. Moreover, the FlowManager is a Ryu controller program that allows the user to manipulate the flow tables in an OpenFlow network manually. We have used the Ryu controller for SDN networking environments due to their ease of deployment, expansion, and simple architecture. Hence, Ryu controllers with OpenFlow-enabled switches are widely used for SDN networking. They are especially suitable for light traffic communication and control. In addition, the Ryu controller provides a routing link to OpenFlow switches to ensure that the topology can perform data analysis. 
Thus, to emulate our network structure, linear topology is used on Mininet, in which 8 switches are connected to the Ryu controller, and each switch is connected to 8 hosts. In total, 64 hosts are linked to the OpenFlow virtual switches, as shown in Figure 8. 
The IP address of the Ryu controller is 192.168.162.133. Likewise, each host is assigned an IP address. For example, the IP address of Host1="10.0.0.1/24" and the mac address starting from 00:00:00:00:00:01 converted from hexadecimal to an integer.
\begin{figure}[h]
\centering
\includegraphics[width=0.5\textwidth]{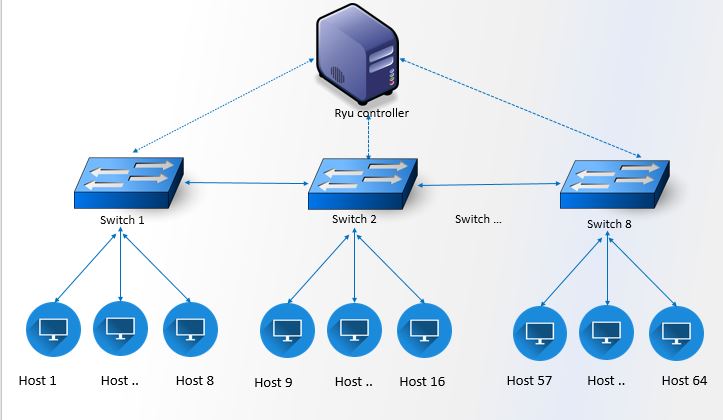}
\caption{Emulated SDN network on scenario 1}
\label{img8}
\end{figure}

In general, the following processes are involved in scenario 1: the data generation and collection process, the detection process, and the mitigation process. These processes are deployed using Mininet VM and Ryu controller VM based on Python programming language.

\begin{itemize}
\item \textit{\textbf{Data generation and collection process:}}
SDN dataset is created using both the Mininet emulator and Ryu controller. The normal traffic is collected using the “iperf” command, and we consider one host (Host1) as a Simple HTTP Server listening on port 80. Additionally, we collect the SYN flood traffic data using the Hping3 tool with random IP addresses. Hping3 is an open-source TCP/IP protocol used as a packet generation tool, that is written in the TCL language. Hping3 enables programmers to create scripts for TCP/IP packet handling and analysis in a restricted period.
MAC addresses are an important criterion to mitigate SYN flood attacks because layer 2 switches forward incoming traffic based on Mac addresses. Also, it helps to identify the infected source port. Moreover, the layer 4 switch depends on the source and the destination ports that are essential in the flow table with the following features: datapath id, source IP, source Mac, destination IP, destination Mac, IP protocol, ICMP code, ICMP type, packet counts, and flags. Table \ref{Table 1} provides detailed information about the collected dataset.
\begin{table}[]
\caption{the collected dataset using SDN}
\begin{tabular*}{200pt}{@{\extracolsep\fill}lcccc@{\extracolsep\fill}}
\hline
\textbf{SDN dataset} & \textbf{All samples} &\textbf{BEGNIN} & \textbf{SYN}  \\ 
\hline
Train/test & 737156 &   816156 &   1553311           \\ \hline
\end{tabular*}
\label{Table 1}
\end{table}
\item\textit{\textbf{The detection process:}}
 After the pre-processing of the collected data presented in \ref{Table 1}, we will split the dataset as follows: The training set contains $80\%$ of the dataset, whereas the testing set contains $20\%$ of the dataset. Then, we use the ANFIS algorithm with cross-validation to avoid overfitting and train the collected dataset to achieve an accuracy of $100\%$. Next, once the packet-in is received in various forms of regular traffic and attack traffic, the Ryu-controller collects the features and assigns their values to the predicted dataset. For the prediction process, the detection module (ANFIS algorithm) examines each flow entry. 

\item\textit{\textbf{The mitigation process:}}
DDoS attacks are difficult to mitigate because of IP spoofing; therefore, blocking the suspected attacker's IP is ineffective in mitigating; To achieve our objective of obtaining a list of edge switches directly connected to each host, we will store the Mac address, port number, and switch ID for each host in a Python dictionary. This dictionary will serve as a data structure to retrieve the required parameters for creating mitigation rules.\\
Every flow entry passes the detection process to check if it is a normal packet or a malicious packet. Then, it will be sent to the Ryu controller to make a decision based on the result of the prediction. Therefore, if the flow entry’s predicted value is 1, it indicates an SYN flood attack in which the attacker transmits both the real source Mac address and a random false source IP, repeating the higher Mac address with different IPs in each flow entry indicates that the hacker is the host of this Mac address. In this case, we use the assigned Mac address to get the port number and switch id from the dictionary. The Ryu controller then responds by enforcing the rule that rejects all packets originating from that attacker, This rule is then sent to the affected switch, instructing it to block the specific port that is directly connected to the attacker's host. By implementing this rule, the switch effectively prevents any communication from the attacker's host through that particular port, helping to mitigate the impact of the attack.
Both the hard timeout and the idle timeout are essential parameters that must be adjusted for the  mitigation process: 
\begin{itemize}
\item Idle time means the flow rule will be deleted if no match occurs with incoming packets within the idle timeout value.
\item Hard timeout means the flow rule will be deleted automatically after hard timeout expires since the rule is created.
\end{itemize}

\begin{table}[hbt!]
\caption{Experiment Parameters}
\begin{tabular*}{200pt}{@{\extracolsep\fill}lcc@{\extracolsep\fill}}
\hline
\textbf{Parameters} & \textbf{Value}   \\ 
\hline
Traffic generation tool        & iperf, Hping3        \\ \hline
Simulation time  & 140 sec       \\ \hline
Bandwidth   & 100 Mbits/sec        \\ \hline
Data collection interval      & 5 sec        \\ \hline
Server   & host 1         \\ \hline
\end{tabular*}
\label{Table 2}
\end{table}
In the case of an attack, the Ryu controller blocks the packet on the OF switch with idle time = 0 sec and hard time = 300 sec with a high priority, we used priority 1000 for our model. As a result, the switch continues to block the source port for 300 seconds without notifying the controller.
 Otherwise, if the detection result is 0, this signifies normal traffic. the idle time will be 200 seconds, and each flow entry has a fixed priority of 10. If no matching happens throughout this time period, the flow rule will be removed after 200 seconds. The hard time will be 400 seconds, after which all flow entries will be deleted.\\
 During this experiment, the real-time flow traffic captured by Wireshark is represented in Figure 9 display the packets per second versus the time plot.  Additionally, Table  \ref{Table 2} presents the parameters employed in this experiment.\\ Initially, normal traffic is sent out at time 0 seconds. Next, a Syn flood attack is initiated, at time 60 the packet rate reaches a threshold value close to 700 packets per second. The ANFIS detection module identifies the attack when  Once the attack is detected, the mitigation module takes over. The controller utilizes appropriate flow rules to mitigate the attack by dropping packets, blocking the source ports involved in the attack, and informing the switches to update the flow table accordingly. The attack is successfully mitigated in less than 5 seconds, resulting in a significant drop in the packet rate. the graph shows the continued  normal traffic flow without any breakdown until the end of the experiment 140 seconds. This period is crucial as it represents  the controller's capability to receive packets effectively.
 Figure 10 can demonstrate that, During the attack, we observed a decrease in bandwidth consumption , reaching as low as 90 Mbits/sec. Fortunately, it quickly recovered to its pre-attack state and remained relatively stable at around 100 Mbits/sec. This demonstrates the effectiveness of our model in mitigating the impact of the attack and restoring normal network performance.

\begin{figure}[h]
\centering
\includegraphics[width=0.5\textwidth]{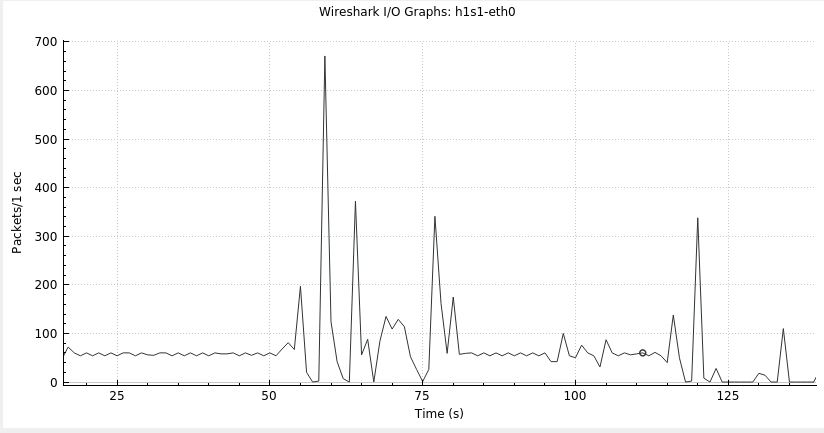}
\caption{real-time Detection and mitigation of Syn flood attack}
\label{img9}
\end{figure}
\begin{figure}[h]
\centering
\includegraphics[width=0.5\textwidth]{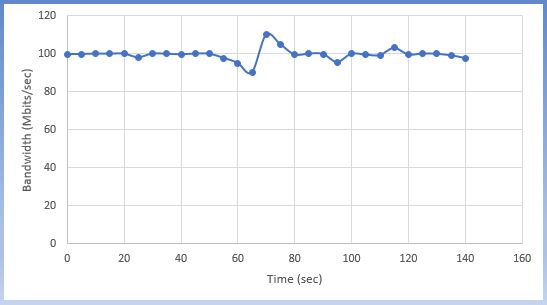}
\caption{Bandwidth usage}
\label{img10}
\end{figure}
\end{itemize}
\item[b.] \textbf{Scenario 2}
 In the second scenario, we evaluate the proposed model's capability to identify the TCP SYN flood DDoS attacks using the CIC-DDoS dataset produced by Sharafaldin et al. (2019) \cite{ar55} for detecting DDoS attacks and classifying attack types. This dataset is in a CSV format, It includes both benign and current popular DDoS attacks launched in 2019. It is collected on the first and second days and reflects the actual real-world data (PCAPs). It also provides the findings of a network traffic analysis performed with CICFlowMeter-V3 that includes labeled traffic flows. This dataset originally had 88 features.

In this scenario, we use the SYN flood dataset presented in Table \ref{Table 3}. TCP SYN flood is a type of exploitation category-based DDoS attack that exploits vulnerabilities in TCP connection protocols. It is composed of data from two days, each with a different attack category and a wide range of imbalance class distribution.

\begin{table}[hbt!]
\caption{SYN flood CIC-DDoS 2019 dataset}
\begin{tabular*}{250pt}{@{\extracolsep\fill}lcccc@{\extracolsep\fill}}
\hline
\textbf{All New dataset}   & \textbf{All samples} & \textbf{BEGNIN} & \textbf{SYN}  \\ 
\hline
Training day & 1582681 &   392 &   1582289             \\ 
\hline
Testing day & 4320541 &   35790 &   4284751             \\ 
\hline
\end{tabular*}
\label{Table 3}
\end{table}
\begin{itemize}
\item\textit{\textbf{Resampling data:}}
  Both training and testing datasets have a minority class "BENIGN" with a little sample, resulting in an imbalanced classification, which has an impact on a model's capacity to learn and decide, furthermore, can cause overfitting in our model. To accomplish this, we build a new dataset in which we take all samples labeled "BENIGN" from the training and testing datasets, forming $10\%$ of the total dataset and $90\%$ of samples labeled "SYN" as shown in Table \ref{Table 4}.
\begin{table}[hbt!]
\caption{the new balanced SYN flood dataset}
\begin{tabular*}{200pt}{@{\extracolsep\fill}lcccc@{\extracolsep\fill}}
\hline
\textbf{New dataset} & \textbf{All samples} &\textbf{BEGNIN} & \textbf{SYN}  \\ 
\hline
Train/test & 180910 &   36182 &   144728           \\ \hline
\end{tabular*}
\label{Table 4}
\end{table}
\item\textit{\textbf{Data Pre-processing:}}
In this section, we will go over the techniques used to analyze our dataset, which contains 88 features. The data will be cleansed and prepared to use in our suggested ANFIS algorithms once certain undesirable attributes have been removed and adjusted. As a result, the implementation of a data preprocessing step, as shown in Figure 11, provides more reliable training and, thus, a more accurate model.

\begin{itemize}
 \item[\checkmark] First, we removed features that have a unique value in the entire dataset that do not affect the training phase ('Bwd PSH Flags',' Fwd URG Flags', ' Bwd URG Flags', 'FIN Flag Count','Fwd Avg Bytes/Bulk',  Fwd Avg Packets/Bulk', ' Fwd Avg Bulk Rate', 'Bwd Avg Bytes/Bulk', ' PSH Flag Count',' ECE Flag Count',' Bwd Avg Packets/Bulk', 'Bwd Avg Bulk Rate').
 \item[\checkmark]  Some values of 'Init Win bytes forward' and 'Init Win bytes backward' of flow data from the Syn csv file were set to -1. Nevertheless, it is inconceivable to initiate a byte window of size -1, this problem was caused by a software issue with CICFlowmeter and should be set to 0 or removed to not disrupt the training phase.
\begin{figure}[h]
\centering
\includegraphics[width=0.5\textwidth]{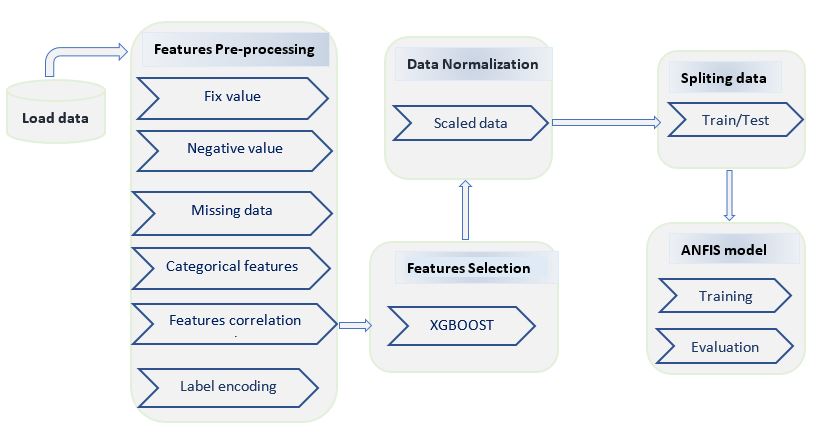}
\caption{Data Pre-processing}
\label{img111}
\end{figure}

\item[\checkmark] The need to cope with missing data threw off the model's training. The lines containing 'infinity' and 'NaN' were removed from 'Flow Bytes/s' and 'Flow Packets/s'.
\item[\checkmark] We removed categorical features that can change from one network to another ('Source Port', 'Destination Port', 'Source IP', 'Destination IP', 'Flow ID', 'SimillarHTTP', 'Unnamed: 0', 'Timestamp'). 
\item[\checkmark] To properly distinguish important features, delete columns with a correlation higher than 0.8 (' Total Backward Packets', ' Total Length of Bwd Packets', ' Fwd Packet Length Std', ' Bwd Packet Length Min', ' Bwd Packet Length Mean', ' Bwd Packet Length Std', ' Flow IAT Mean', ' Flow IAT Std', ' Flow IAT Max', 'Fwd IAT Total', ' Fwd IAT Mean', ' Fwd IAT Std', ' Fwd IAT Max', ' Fwd IAT Min', ' Bwd IAT Std', ' Bwd IAT Max', ' Fwd Header Length', ' Bwd Header Length', ' Max Packet Length', ' Packet Length Mean', ' Packet Length Std', ' Packet Length Variance', ' RST Flag Count', ' Average Packet Size', ' Avg Fwd Segment Size', ' Avg Bwd Segment Size',' Fwd Header Length.1', 'Subflow Fwd Packets', ' Subflow Fwd Bytes', ' Subflow Bwd Packets', ' Subflow Bwd Bytes', ' Active Max', ' Active Min', 'Idle Mean', ' Idle Max', ' Idle Min').
\item[\checkmark] In order to detect and classify DDoS attacks, the dataset is split into two classes. The label "BENIGN" is coded as "0" and the label "Syn" is coded as "1" in the dataset created to detect a SYN flood DDoS attack on the network traffic.  
\item[\checkmark]Feature selection is used to discover key data features and decrease the amount of data required for detection. we use the XGBoost technique that provides an importance score to each feature based on its influence in making crucial decisions using boosted decision trees \cite{ar57}. Then, depending on the rated feature, we removed features that were of negligible importance ' Protocol', ' Flow Duration',' Total Fwd Packets',  ' Fwd Packet Length Max',   'Bwd Packet Length Max', ' Flow IAT Mean', ' Flow IAT Min','Bwd IAT Total', ' Bwd IAT Mean', ' Bwd IAT Min', 'Fwd PSH Flags','Fwd Packets/s', ' Bwd Packets/s', ' Min Packet Length',' SYN Flag Count',' CWE Flag Count', ' Down/Up Ratio',' Init Win bytes backward', ' act data pkt fwd', 'Active Mean', ' Active Std', ' Idle Std', and choose nine ideal feature subsets, as presented in Table \ref{Table 5}.\\

\begin{table*}[!htb]%
\centering
\caption{Features selected with XGBoost}
\begin{tabular*}{350pt}{@{\extracolsep\fill}lcccc@{\extracolsep\fill}}

\hline
\textbf{Feature Name}   & \textbf{Description}  \\ 
\hline
\textit{Total Length of Fwd Packets} & Overall size of packet in the forward direction.                 \\ 
\hline
\textit{Fwd Packet Length Mean}  & Mean size of packet in forward direction.     \\ 
\hline
\textit{ACK Flag Count} & Number of packets with ACK.      \\ 
\hline
\textit{URG Flag Count}  & Number of packets with URG.      \\ 
\hline
\textit{Init Win bytes forward} & Number of bytes sent in initial window in the forward direction. \\ 
\hline
\textit{min seg size forward} & The observed minimum segment size in the forward direction.     \\ 
\hline
\textit{Inbound} & The direction in which traffic moves between networks. \\
\hline
\textit{Label} & Type of packets for classification.    \\ \hline
\end{tabular*}
\label{Table 5}
\end{table*}

\item We normalize the data by scaling all features in the range of 0–1 value.  As previously described, The dataset was divided into two parts training data and testing data.  by using cross-validation to avoid overfitting in training steps. 
\item Finally, we put the ANFIS model to the test for making predictions on unseen data. The next section discusses the performances and results.
\end{itemize}
\end{itemize}
\end{enumerate}
\subsection{Performance metrics}
Using the right performance metrics is the key to correctly evaluating models. Therefore, in this section, we explore the following performance metrics to evaluate the FASA framework:
\begin{itemize}
\item \textit{True Negatives (TN)}: Normal flow data is appropriately identified as such.
\item \textit{True Positives (TP)}: malicious flow data is accurately identified as such.
\item \textit{False Positives (FP)}: Normal flow data is mistakenly labeled as malicious traffic.
\item \textit{False Negatives (FN)}: malicious flow data is classed as normal flow data when it isn't.
\end{itemize}

In addition, we provide the confusion matrix to describe our model's classification performance. It can resume the correct and false predictions obtained using our proposed approach, as demonstrated in Figure 12.
\begin{figure}[ht]
\begin{subfigure}{.5\textwidth}
  \centering
  \includegraphics[width=.7\linewidth]{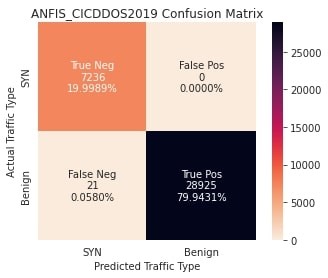}  
  \caption{Using CIC-DDoS dataset}
  \label{fig:sub-first}
\end{subfigure}
\begin{subfigure}{.5\textwidth}
  \centering
  \includegraphics[width=.7\linewidth]{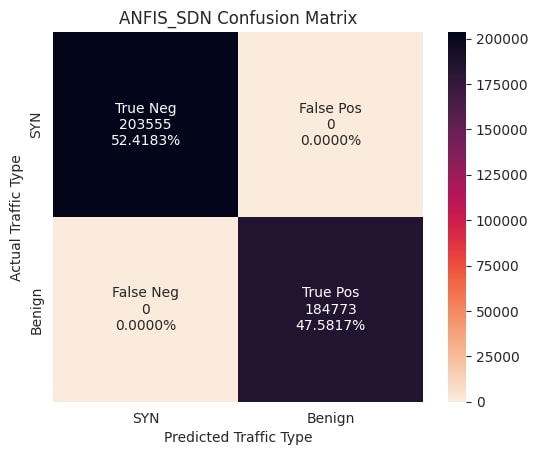}  
  \caption{Using SDN dataset}
  \label{fig:sub-second}
\end{subfigure}
\caption{Confusion matrix of the ANFIS model}
\label{fig:fig}
\end{figure}

 Accurately distinguishing the Benign class within our model is of utmost importance, as elevated false positive rates can result in unnecessary complexity and unwarranted alerts. Our main objective is to minimize the false rate. Hence, our framework achieves a rate of $0\%$ of false positives in both CIC-DDoS2019 and SDN datasets. Otherwise, it obtains $0.058\%$ false negatives in the CIC-DDoS2019 dataset and $0\%$ in the SDN dataset.
the Receiver Operating Characteristic (ROC) curve is performed. It represents the relation between both the True and False parameters. The area under the ROC Curve (AUC) measures whether it is possible to distinguish false positives from true positives. As illustrated in Figure 13, our model has an AUC of $99.96\%$ using the CIC-DDoS2019 dataset and $100\%$ using the SDN dataset and there are two extremely similar values, indicating that our suggested model separates correctly positive from negative classes. By employing established techniques like k-fold cross-validation, the model ensures generalizability and guards against overfitting. Furthermore, the meticulous selection and optimization of impactful traffic features enhance the model's proficiency in distinguishing between normal and attack behaviors. Additionally, the fusion of fuzzy logic and neural learning components proves effective in capturing complex traffic patterns. Lastly, training on diverse attack data distributions further enhances the model's robustness.\\
\begin{figure}[h]
\centering
\includegraphics[width=0.5\textwidth]{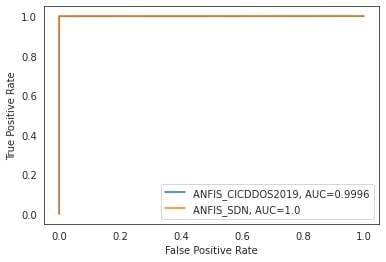}
\caption{ROC Cuvre of ANFIS model}
\label{img13}
\end{figure}
We have also used a variety of measures to assess our suggested model, including accuracy, precision, recall, and F-score, to conduct an in-depth comparative assessment with some other relevant methods. These metrics, which are often employed in SYN flood DDoS detection systems, are described in the following:
\begin{enumerate}[1.]
 \item Accuracy refers to the ratio of the number of samples correctly classified to the overall number of samples observed. It is computed as follows:
\begin{equation}
accuracy=\frac{tp+tn}{tp+tn+fp+fn} 
\end{equation}
 \item The precision is the ratio of correctly predicted positive samples, it is calculated as follows:
\begin{equation}
precision=\frac{t p}{t p+f p}
\end{equation}						
 \item The false positive rate is determined by calculating the proportion of negative samples that were incorrectly classified as positive using the following formula:
 \begin{equation}
 fp-rate= \frac{f p}{f p+t n}	
\end{equation}
 \item The recall also called the true positive rate, is calculated with the ratio of correctly discovered positive samples, It is determined using the equation:
\begin{equation}
recall=tp-rate= \frac{t p}{t p+f n}   \end{equation}
\item Good precision may be more relevant in certain situations, whereas high recall might be more critical in others. Across many cases, though, we aim to enhance both values. The f1-score is the combination of these values, and it is commonly stated as the harmonic mean:
 \begin{equation}
 f1-score= \frac{2\times precision\times recall}{(precision+recall)}
 \end{equation}
\end{enumerate}
\subsection{Evaluation results}
To validate our system, we have compared the FASA framework to the FUPE \cite{ar59} method and other DDoS attack detection systems that were employed on SDN and used the CIC-DDoS 2019 dataset, as illustrated in Table \ref{Table 6}. 

\begin{table}[hbt!]
\caption{The evaluated metrics were used to compare the results of ANFIS with other methods.}

\begin{tabular*}{250pt}{@{\extracolsep\fill}lccccc@{\extracolsep\fill}}

\hline
\textbf{Method}             & \textbf{Accuracy} & \textbf{precision} & \textbf{Recall} & \textbf{F1-score} \\ \hline
\textbf{ANFIS   SDN}        & 100               & 100                & 100             & 100               \\ \hline
\textbf{ANFIS CIC-DDoS2019} & 99.95             & 100                & 99.94           & 99.95             \\ \hline
\textbf{FUPE \cite{ar59}} & 98.2             & 96.08                & 98          &  N/A           \\ \hline
\textbf{CNN \cite{ar24}}       & 95.4              & 93.3               & 92.4            & 92.8              \\ \hline
\textbf{GAN \cite{ar25}}       & 94.38             & 94.08              & 97.89           & 95.94             \\ \hline
\textbf{MLP \cite{ar26}}       & 95.01             & 95.46              & 94.51           & 94.98             \\ \hline
\end{tabular*}
\label{Table 6}
\end{table}
The first method is FUPE \cite{ar59} that puts forward a fuzzy-based multi-objective particle swarm Optimization approach, a security-aware task scheduler in IoT–fog networks. The second method is the Convolutional Neural Network (CNN) \cite{ar24}, a low-cost based supervised classifier designed to identify suspicious events in a data center. the next approach is based on Generative Adversarial Network GAN \cite{ar25} for identifying DDoS threats in SDN environments. Finally, the Multi-layer Perceptron (MLP) \cite{ar26} is adopted to identify and prevent Low Rate-DDoS attacks in SDN settings. 
Figure 14 depicts a comprehensive analysis of the metric findings of the comparative approaches.

\begin{figure}[h]
\centering
\includegraphics[width=0.5\textwidth]{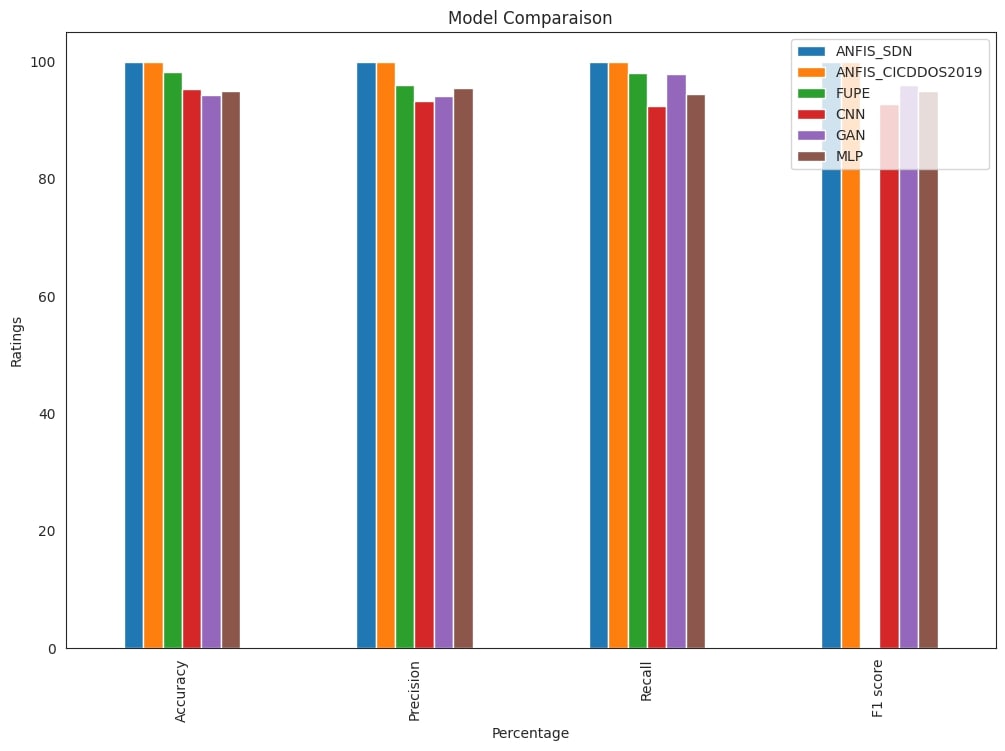}
\caption{ Evaluation metrics for comparative methods }
\label{img14}
\end{figure}

As shown in Figure 14, we can observe that the performance of our model using the SDN dataset outperforms all previous techniques, with $100\%$ accuracy, precision, recall, and F1-score in each case and it closely resembles the outcome obtained using the CIC-DDoS2019 dataset. In addition, the accuracy of every learning algorithm is assessed. As a result, The ANFIS achieved the highest accuracy rating of $99.95\%$ across all classifiers, then the FUPE approach with $98.2\%$ followed by the CNN algorithm with $94.83\%$. Furthermore, MLP and GAN classifiers attained an accuracy of $95.4\%$ and $95.01\%$, respectively. It also illustrates the precision of each algorithm in identifying legal and malicious traffic. Thus, the ANFIS reached $100\%$ precision, and FUPE with a precision of $96.08\%$, and the MLP attained a precision value of $95.46\%$. Next, the GAN, and CNN algorithms with a precision of $94.08\%$, and $93.3\%$, respectively.
Furthermore, Figure 14 displays the recall values of all methods used in the performance evaluation. The ANFIS algorithm had a $99.94\%$ recall value followed by FUPE with $98\%$, whereas GAN had a $97.89\%$ recall rating. In comparison to the other algorithms tested, the CNN achieved the lowest recall value of $92.4\%$ while the MLP had a recall of $94.51\%$.
It also illustrates the F1-score of the classifying methods with $99.95\%$, the ANFIS received the highest F1-Score. On the other hand, GAN, MLP, and CNN received F1-scores of $95.94\%$, $94.98\%$, and $92.8\%$, respectively. While the FUPE's F1-Score is not mentioned.
In conclusion, our FASA framework outperforms the other evaluated approaches. The promising test results indicate that it is an effective approach for identifying SYN flood DDoS attacks.
\section{Conclusion and future work} 
\vspace{-1ex}
In this work, FASA, a Fog computing-based SYN Flood DDoS attacks mitigation using an Adaptive Neuro-Fuzzy Inference System (ANFIS) and Software Defined Networking (SDN) Assistance was proposed. The choice of the integration of SDN and fog environment with the ANFIS machine learning algorithm brings intelligence to the SDN controller. Also, it makes our framework suitable, efficient, and more secure against SYN flood attacks. We trained and evaluated our framework on the newly released CIC-DDoS2019 dataset that contains the most recent and extensive SYN flood DDoS attacks. The findings of the performance assessment indicate that the suggested model has a high detection accuracy and a low rate of false positive and negative rates, which is a remarkable result and it also offers the highest evaluation metrics regards to precision, recall, and F-score when compared to well-known machine learning algorithms.
Our future work is to focus on how well our proposed model performs on various datasets. In the current experiments, we have employed a binary classification approach that is implemented on SDN to distinguish between legitimate and malicious input traffic in fog computing. Thus, in future work, we will try to investigate the utility of the suggested approach for other multi-class classification systems. Additionally, to create a diversified dataset that truly represents actual internet traffic, we will emulate the SDN network under various scenarios and with various attack traffic. In addition, we will also consider expanding our work to include the SoDIP6-based ISP/Telecom network, including edge computing network scenarios. This will allow us to evaluate the performance of our proposed model in a more complex and realistic environment. We will also investigate the use of our model for other network security applications, such as intrusion detection and prevention.\\

\section*{Conflicts of Interest}
All authors declare no conflict of interest.

\section*{Acknowledgments}
The authors conducted this research while affiliated with Abou Bekr Belkaid Tlemcen University, Paris-Saclay University, Edinburgh Napier University, and  Dakahlia Mansoura University.

\nocite{*}
\bibliography{bibfile} 

\balance
\bibliographystyle{IEEEtran}


\end{document}